\documentclass[aps,letterpaper,amssymb,twocolumn,pra]{revtex4}

\usepackage{bm}
\usepackage{amssymb}
\usepackage{times}
\usepackage{latexsym}
\usepackage{graphicx}
\usepackage{color}
\def\nm{{\ {\rm nm}}}                       

\def\kHz{{\ {\rm kHz}}}                     

\def\us{{\ \mu{\rm s}}}                     
\def\ms{{\ {\rm ms}}}                       

\def\Er{E_r}                            
\def\Rb87{^{87}\text{Rb}}                     
\def\Na23{^{23}\text{Na}}                     
\def\Cs133{^{133}\text{Cs}}                     
\def\Li6{^{6}\text{Li}}                       
\def\K40{^{40}\text{K}}                       

\def\ket#1{\mathinner{|{#1}\rangle}}

{\catcode`\|=\active
  \gdef\Braket#1{\left<\mathcode`\|"8000\let|\BraVert {#1}\right>}}
\def\BraVert{\egroup\,\mid@vertical\,\bgroup}

\begin{document}
\title{Chern numbers hiding in time-of-flight images}
\author{Erhai Zhao$^{1,2}$, Noah Bray-Ali$^2$, Carl J. Williams$^2$, I.~B.~Spielman$^2$ and Indubala I. Satija$^{1,2}$}
\affiliation{$^{1}$School of Physics, Astronomy, and Computational Sciences, George Mason University,
 Fairfax, VA 22030}
\affiliation{ $^{2}$Joint Quantum Institute, National Institute of Standards and
Technology and University of Maryland, Gaithersburg, MD 20899}
\date{\today}

\begin{abstract}
We present a technique for detecting topological invariants -- Chern numbers -- from 
time-of-flight images of ultra-cold atoms. 
We show that the Chern numbers of integer quantum Hall states of lattice fermions
leave their fingerprints in the atoms' momentum distribution.
We analytically demonstrate that the number of local maxima in the momentum distribution
is equal to the Chern number in two limiting cases, for
large hopping anisotropy and in the continuum limit. 
In addition, our numerical simulations beyond these two limits show that these local maxima
 persist for a range of parameters.
Thus, an everyday observable
in cold atom experiments can serve as 
a useful tool to characterize and visualize quantum states with non-trivial topology.
\end{abstract}
\maketitle

\section{introduction}
Topological insulators (TI) and superfluids are many-body quantum systems with energy gaps in the bulk but topologically protected, gapless excitations on the boundary~\cite{hasan10,qi10}. Each class of topological phases is 
characterized by a topological invariant which can be either an integer or a binary quantity. For example, the
integer quantum Hall insulators~\cite{klitzing80} are distinguished from other two-dimensional insulators with
the same symmetry by the Chern number, an integer that capture the global properties of all occupied bands ~\cite{thouless82}. The Chern number coincides with the number of chiral edge modes, and the Hall conductance in units of the conductance quantum. 
A central problem in exploring new forms of topological matter
is how to extract the topological invariants from experiments, and unambiguously determine
the topological phase.
   
Experiments with ultracold atoms has brought us opportunities to 
creating topological phases of matter in parameter regimes unreachable in solids. 
In addition to band-structure engineering using optical lattices and tuning the atom-atom 
interaction by Feshbach resonance, artificial magnetic fields for neutral atoms have been produced
by either rotation~\cite{cooper-rev} or atom-laser coupling~\cite{spielman09,dalibard10}, 
making way to realizing quantum Hall states of cold atoms~\cite{umucallar08}.
Moreover, non-Abelian synthetic gauge fields~\cite{goldman10,dalibard10} and spin-orbit coupling~\cite{lin11} 
have been demonstrated. These developments motivated theories on
properties of ultracold topological matter, e.g.,
the fractional quantum Hall effect of interacting bosons~\cite{cooper-rev,demler-etal,Palmer2008}, 
the quantum spin Hall effect of fermions~\cite{goldman10}, and the 
quantum anomalous Hall states of $p$-orbital fermions~\cite{zhang11}.

While cold atom systems offer new detection techniques and great tunability, 
it is challenging to measure the topological invariants directly. Measurements
of local density of states (of the edge states) or mass transport 
are marred with complexity in contrast to solid state systems. Thus, creative methods
have to be invented. For example, Ref. \cite{umucallar08} proposed to 
extract the quantized Hall conductance from the derivatives of the in-situ atom density 
distribution by the St\v{r}eda formula. Here, we show that the Chern number $c_r$ of 
a lattice-quantum Hall system of fermions can be simply determined from the
structures (ripples) in the momentum distribution $n({\bf k})$ as measured in 
time-of-flight (TOF) images for a range of parameters.
Our results transform $n({\bf k})$, a simple everyday observable 
of cold-atom experiments, into an unexpected, useful tool for characterizing and
visualizing topological states of matter. 

\begin{figure}
\includegraphics[width=2.8in]{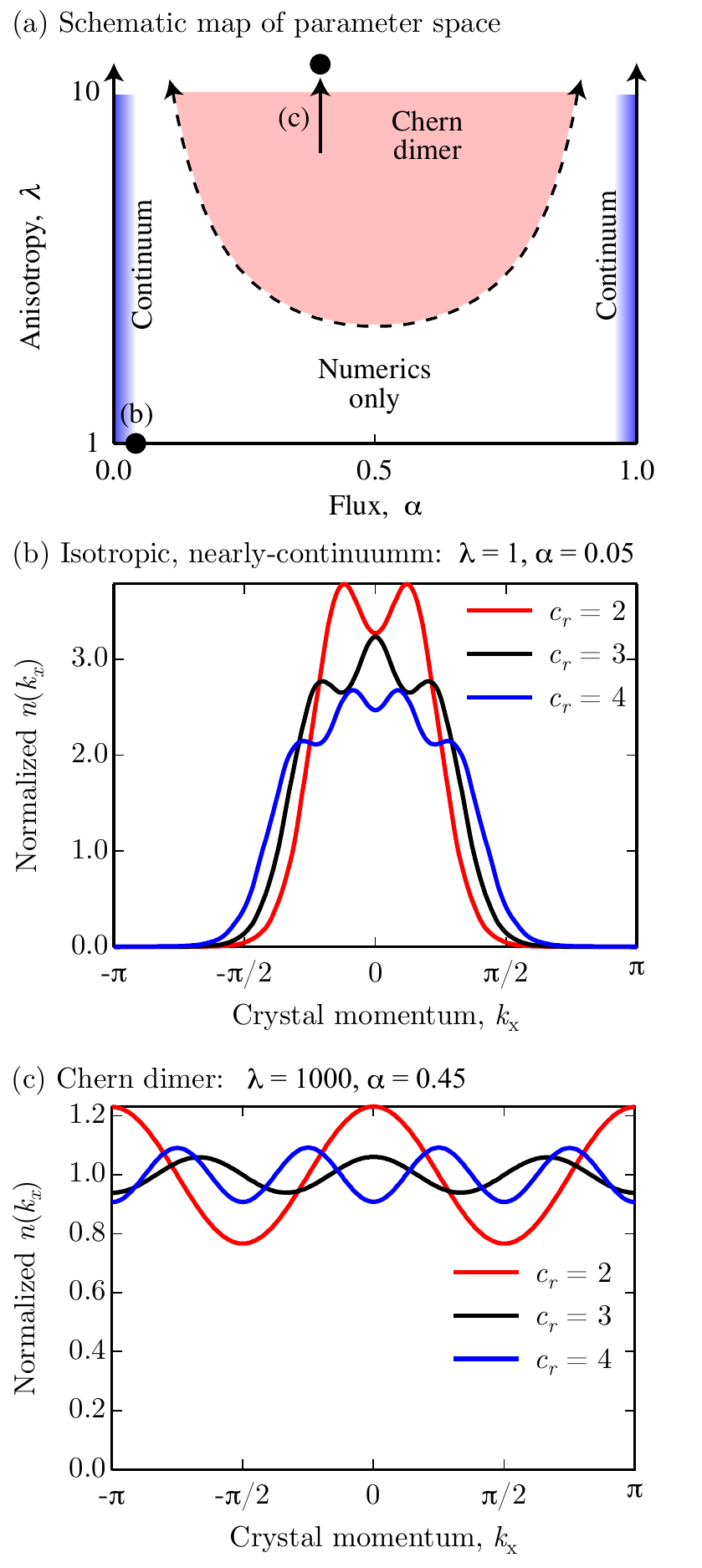}
\leavevmode
\caption{(Color) Relation between the Chern number $c_r$ and structure in TOF data as a function of magnetic flux per unit cell $\alpha$ in units of the flux quantum $h/e$, and hopping anisotropy $\lambda$.  (a) Schematic map of the $\alpha$-$\lambda$ plane showing the Chern-dimer and the continuum regimes. Numerical results for the remainder of parameter space will be shown in Fig. \ref{F2} and \ref{F3}.  (b,c) Normalized 1D crystal momentum distributions $n(k_x)$, see Eq.~(\ref{MomentumX}), computed at three different chemical potentials corresponding to $c_r = 2,3,4$; in each case, the number of peaks or oscillations is equal to $c_r$.  The data in (b) is for an isotropic $\lambda=1$ lattice at $\alpha=0.05$, and the data in (c) is for an anisotropic $\lambda = 1000$ lattice at $\alpha = 0.45$. 
$k_x$ is measured in units of $1/d$, where $d$ is the lattice spacing.
} 
\label{F1}
\end{figure}

Figure~\ref{F1}a summarizes the main results of this paper obtained for the Hofstadter model~\cite{hofstadter76}.
We highlight the parameter regimes where the structure in $n({\bf k})$ is most evident: 
in the continuum limit (Fig.~\ref{F1}b) or for highly anisotropic lattices (Fig.~\ref{F1}c).
In particular, we provide a transparent understanding of $n({\bf k})$
for large hopping anisotropy. In this limit, 
states with energy at the band edges dominate, and they are dimerized (their localized wave functions are 
peaked at pairs of spatially separated lattice sites). The size of these ``Chern-dimers,'' in units of the lattice spacing $d$, is equal to the Chern number of the corresponding gap. 
As a result,
$n({\bf k})$ varies sinusoidally (Fig.~\ref{F1}c) with a period determined by the Chern number.
Furthermore, our detailed numerical studies verify that such distinctive features in the 
momentum distribution persist beyond the Chern-dimer and the continuum limit
(Fig.~\ref{F2}c and \ref{F2}d).
However, the correspondence between the number of local maxima in $n({\bf k})$ and the Chern number 
is not always precise for arbitrary parameters.

\section{Ultracold fermions in synthetic magnetic field}
We consider the Hofstadter model~\cite{hofstadter76} of spinless fermionic atoms on a square optical lattice in the ${\bf e}_x$-${\bf e}_y$ plane 
\begin{equation}
H=-\sum_{\langle{\bf s}, {\bf s'}\rangle} t_{{\bf s}, {\bf s'}} \left (e^{i\theta_{{\bf s}, {\bf s'}}}f^\dagger_{\bf s} f_{\bf s'} + {\rm h.c.} \right)- \sum_{\bf s} \mu_s f^\dagger_{\bf s} f_{\bf s}. \label{eq-hof}
\end{equation}
Here, $f^\dagger_{\bf s}$ is the creation operator for a fermion at site ${\bf s}=\left\{s_x,s_y\right\}$; the
nearest-neighbor hopping strengths $t_x = t_{{\bf s},{\bf s}+{\bf e}_x}$, and $t_y = t_{{\bf s},{\bf s}+{\bf e}_y}$ are real and positive; $\theta_{{\bf s,s'}}= e \int_{\bf s}^{\bf s'}{\bf A}\cdot d{\bf l}$ depends on the vector potential ${\bf A}$~\footnote{In cold atom experiments the product $e{\bf A}$ enters the Hamiltonian, and neither $e$ nor ${\bf A}$ is separately defined.} for particles with charge $e$; and $\mu_{\bf s}$ is a local chemical potential.  With cold atoms, the hopping anisotropy  $\lambda=t_y/t_x$ can be tuned by varying the intensity of the lasers
giving rise to the optical lattice potential~\cite{spielman09}.  For an applied magnetic field ${\bf B} = B {\bf e}_z$ normal to the 2D plane, 
$\alpha = d^2 e B/h$ is the magnetic flux per plaquette in units of the flux quantum.
We study homogeneous gases with $\mu_{\bf s}=\mu$, and also finite systems in the presence of an isotropic harmonic trap potential, captured by $\mu_{\bf s}=\mu_0-m\omega^2 d^2\mathbf{s}^2/2$.

\begin{figure}
{\includegraphics[width =3.3in]{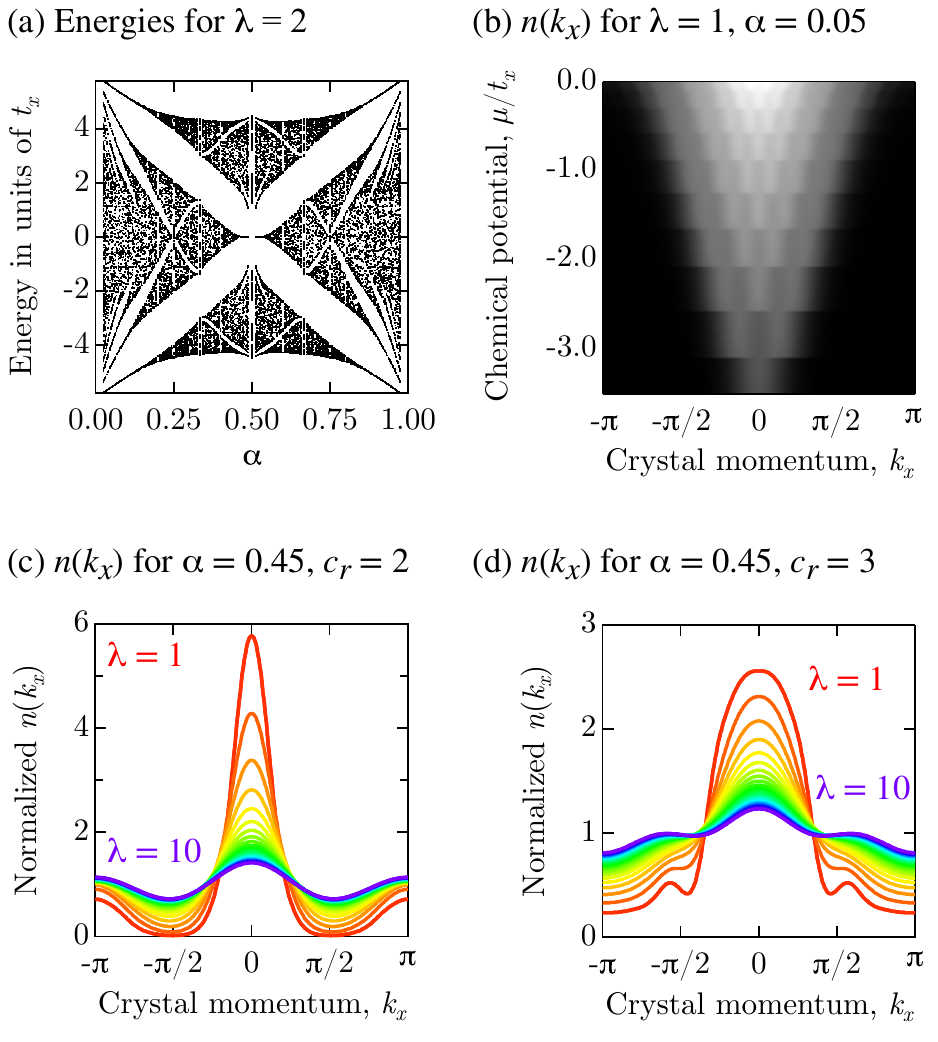}}
\leavevmode
\caption{(Color) 1D crystal momentum distributions $n(k_x)$.  (a) Energy spectrum of Eqs.~(\ref{eq-hof}) and (\ref{harper}) for $\lambda=2$. (b) Evolution of $n(k_x)$ with increasing chemical potential $\mu$ at $\lambda=1$ and $\alpha=0.05$ showing $c_r$ increasing in tandem with the number of oscillations in $n(k_x)$.  (c,d) Evolution of $n(k_x)$ with increasing anisotropy $\lambda$ at $\alpha = 0.45$ for (c) $c_r=2$ and (d) $c_r=3$.  These curves depict the distributions approaching the cosine function with increasing $\lambda$. The unit for $k_x$ is $1/d$.}
\label{F2}
\end{figure}

Artificial magnetic fields have been implemented for bosons~\cite{Lin2009b}, and it is straightforward to apply these techniques to fermions.  With their smaller fine-structure splittings, the alkali fermions ($\Li6$ and $\K40$) are more susceptible to heating from spontaneous emission than some bosons ($\Rb87$ and $\Cs133$), but for the $\hbar/t_{x,y}\sim240\us$ timescales~\footnote{Estimated assuming a $4\Er$ deep lattice from retro-reflected $\lambda_L = 2d =800\nm$ lasers, giving $t_{x,y}=0.085\Er$, where the single photon recoil energy is $\Er = h^2/2 m \lambda_L^2 = h\times7.8\kHz$ and $m$ is the atomic mass.} associated with the major energy gaps in the spectrum (see Fig.~\ref{F2}a), the expected $\approx200\ms$ lifetime in $\K40$ should be acceptable; see Ref.~\cite{goldman10} for an alternate solution.

We first summarize the well known properties of the homogeneous Hofstadter model [Eq. (\ref{eq-hof})]
with commensurate flux $\alpha=p/q$, where the integers $p,q$ are relatively prime. 
In the Landau gauge~\footnote{Our analysis can be generalized to the symmetric gauge which is appropriate for rotating gases, see e.g. S. Powell et al, Phys. Rev. A 83, 013612 (2011).} suitable for current experiments~\cite{spielman09}, the vector potential $\mathbf{A}=  \alpha x {\bf e}_y$ scales the unit cell along ${\bf e}_x$ by a factor of $q$ to a magnetic unit cell; this also creates a magnetic Brillouin zone (MBZ) with $ k_x \in (-\pi/q ,\pi/q]$ 
and $  k_y \in (-\pi,\pi]$. 
 The Hofstadter model is diagonal in the single-particle basis  
 $\Psi_{s_x,s_y}= e^{i(k_x s_x+k_y s_y)} \psi_{s_x}(k_y)$, provided $\psi_{s_x}$ satisfies~\cite{hofstadter76}
\begin{equation}
 e^{ik_x}\psi^r_{s_x+1}+ e^{-ik_x}\psi^r_{s_x-1} + 2\lambda \cos ( 2 \pi s_x \alpha+ k_y)\psi^r_{s_x} = -E_r \psi^r_{s_x}
\label{harper}
\end{equation}
with energy in units of $t_x$.  The index $r=1, 2, ...q$ labels linearly independent solutions, and $\psi^r_{s_x+q} = \psi^r_{s_x}$. 
The single energy band at $\alpha=0$ splits 
into $q$ sub-bands, and the energy spectrum has $q-1$ gaps~\cite{hofstadter76}.
When the chemical potential $\mu$ is inside the $r$-th energy gap, the system
is an integer quantum Hall state with Chern number
$c_r$, defined by the Thouless-Kohmoto-Nightingale-Nijs (TKNN) formula~\cite{thouless82} and related to the quantized Hall
resistance.

Usually we consider TOF images as approximate representations of the momentum distribution $n(\mathbf{k})$ obtained by imaging the 2D atomic column density after a $t_{\rm TOF}$ period of free expansion.  In most cases, the final spatial coordinate of an atom is $\mathbf{x} = \mathbf{v} t_{\rm{TOF}}$ where ${\bf v}$ is the velocity~\cite{gerbier07}, however, vector potentials complicate the interpretation of TOF images because the velocity is ${\bf v}=\hbar {\bf k}/m - e {\bf A}/m$.  In Ref.~\cite{Lin2011a}, we showed that suddenly turning the gauge field to zero is equivalent to a brief electric field that takes the initial momentum $\hbar {\bf k}$ and maps it to a new final velocity ${\bf v'}=\hbar {\bf  k}/m$.   As a result, detected TOF images do approximate momentum distributions~\footnote{In this context, $n(\hbar\mathbf{k})$ is the {\it canonical} momentum distribution, not the mechanical momentum distribution.}.  The crystal momentum distribution
\begin{equation}
\tilde{n}(\mathbf{k})=\sum_{\mathbf{s,s'}}\langle f^\dagger_{\mathbf{s}} f_{\mathbf{s'}}\rangle e^{i\mathbf{k}\cdot(\mathbf{s-s'})},\label{MomentumDistribution}
\end{equation}
can be reconstructed from TOF images using the recipe in Ref.~\cite{Spielman2007}.  Here, the average $\langle\cdots\rangle$ is over the ground state.  We also define a 1D crystal momentum distribution by integrating $\tilde n({\bf k})$ along ${\bf e}_y$, 
\begin{equation}
n(k_x)=\int_{-\pi}^{\pi} d k_y \tilde{n}({\bf k}).\label{MomentumX}
\end{equation}

For homogeneous systems, we computed $n(k_x)$ by numerically diagonalizing Eq.~(\ref{harper}) using
a sufficiently large unit cell, while for finite trapped systems we diagonalized
Eq.~(\ref{eq-hof}).  Our numerical results are complemented by analytic analysis
for low flux and large anisotropy.
In what follows, we discuss in turn the behaviors of $n(k_x)$ in different parameter regimes, 
starting with an isotropic lattice and low flux.

\section{Oscillations in momentum distribution}

Figures~\ref{F1}b and~\ref{F2}b show $n(k_x)$ in this isotropic low flux limit, at $\alpha=0.05$ and $\lambda=1$.  As the chemical potential is increased up to $\mu=0$, successively higher Hofstadter bands are filled, with Chern numbers $c_r=r$ where $r=1,2,3$... labels the highest filled band.  Correspondingly, $n(k_x)$ has exactly $c_r$ local peaks on top of a smooth envelope, i.e., counting peaks in TOF images unambiguously determines the Chern number, and therefore which quantum Hall state the system is in.  To understand this, we recall that  for $\alpha\ll 1$, the cyclotron orbit is large
compared to the lattice spacing, so the system resembles a free Fermi gas in magnetic
field, and the narrow bands in the Hofstadter spectrum map to Landau levels.
For $r$ filled Landau levels, $n(k_x)$ is the summed square modulus of the momentum-space 
wave functions,
\begin{equation}
n(k_x)\propto\sum_{\nu=0}^{r-1} (2^\nu \nu !)^{-1}H^2_\nu( {k_x}{\ell_B})e^{-(k_x\ell_B)^2}, \label{landau}
\end{equation}
where $H_\nu$ are the Hermite polynomials, and $\ell_B=\sqrt{\hbar/eB}$ is the magnetic length. 
Due to the node structure of $H_\nu$, $n(k_x)$ has exactly $r$ local maxima
located at the zeros of $H_{r+1}$.  This is reflected in our numerical results as plotted in Fig.~\ref{F1}b
for the lattice system.  As is depicted by the shaded regions in Fig.~\ref{F1}a, this association is valid in the entire low flux regime and is insensitive to $\lambda$.

At larger $\alpha$ the Landau level picture breaks down, but the correspondence between the number of local peaks and the Chern number appears to persist for a range of parameters.
Fig.~\ref{F2}c and ~\ref{F2}d show examples of $n(k_x)$ for large flux $\alpha=0.45$
and $\lambda=1$ where well resolved peaks do remain. 
More generally, as $\alpha$ increases (for $\mu$ residing within the same energy gap), the distinctive features of $n(k_x)$ are generally retained, but compared to the low flux regime, the local peaks (except at $k_x=0$) move to higher momenta and can be subsumed into the background structure. In some instances, 
these peaks turn into shoulder-like structures.

Intriguingly, as the hopping anisotropy becomes large,
small wiggles on a background turn into
regular and pronounced oscillations
regardless of $\alpha$. Fig.~\ref{F2}c and ~\ref{F2}d
illustrate the progression of $n(k_x)$ from
$\lambda=1$ to $\lambda=10$, showing pronounced oscillations develop with increasing $\lambda$. 
Most importantly, the simple rule reappears in the limit of $\lambda\gg 1$:
the number of oscillations in $n(k_x)$, i.e., the number of peaks for $k_x\in [-\pi,\pi)$, is again exactly
equal to the Chern number $c_r$.  As is emphasized in Fig.~\ref{F1}c, $n(k_x)$ asymptotically approaches 
\begin{equation}
n(k_x)\propto r\pm\frac{1}{2}\cos(c_r k_x) 
\label{nkx-asymp}
\end{equation}
for chemical potential inside the $r$-th gap.  The one-to-one correspondence, or mapping, between the number of oscillations and the Chern number for anisotropic lattices (Fig.~\ref{F1}a: Chern-dimer regime) is one of our key observations which we now explain analytically.

\section{Chern-dimers}
\begin{figure}[b]
{\includegraphics[width =3.3in]{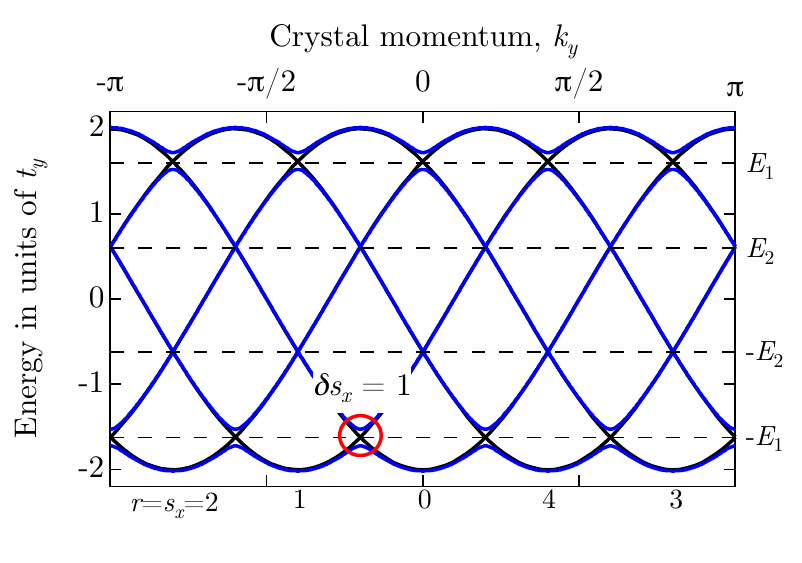}}
\leavevmode
\caption{(Color) 
Energies in the anisotropic limit for $p=1$, $q=5$ showing the perfect sinusoidal energies when $\lambda\rightarrow\infty$ (solid black curves), and the avoided crossings when $\lambda=10$ (solid blue curves).  The hybridized states at these crossings correspond to ``Chern-dimers,'' and the red-circled example has spatial extent $\delta s_x = 1$.
The unit for $k_y$ is $1/d$.
}
\label{F3}
\end{figure}

The $\lambda \gg 1$ limit is tractable in perturbation theory~\cite{fradkin91} where the solutions to Eq.~(\ref{harper})
are simple.  When $t_x = 0$ (i.e., $\lambda\rightarrow\infty$), there are $q$ bands each with dispersion $E_r(k_y)= -2 t_y \cos( 2\pi \alpha r+k_y)$,
$r\in\left\{1,\ldots, q\right\}$ (see Fig.~\ref{F3}, where $p=1$ and $q=5$).  These eigenstates are extended along ${\bf e}_y$ and localized to single sites as a function of $x$.  As defined in Eq.~(\ref{harper}), the $r$-th band describes states $\psi^r_{s_x}=\delta_{s_x,r}$ localized at site $s_x=r$ and displaced in $k_y$, as identified in Fig.~\ref{F3}.  The bands intersect at numerous degeneracy points ${k^*_y}$ for states $r=s_x$ and $r^\prime=s_x^\prime$.  The spatial separation along ${\bf e}_x$ of these localized states is $\ell = s_x^\prime-s_x$.  These band-crossings occur at $q-1$ discrete energies  (right axes of Fig.~\ref{F3}) with exactly $q$ crossings having the same $\ell$ at each energy.  For finite $\lambda$, $t_x$ hybridizes the states at each crossing, leading to $q(q-1)$ avoided crossings, and $(q-1)$ gaps in the band structure.  Near each degeneracy point the system is well described by the effective Hamiltonian~\cite{fradkin91}, 
\begin{equation}
\check{H}_{2}(\mathbf{k})=\Delta_{\ell}[\cos(k_x\ell)\check{\sigma}_x+\sin(k_x\ell)\check{\sigma}_y]
+v_{\ell}(k_y-k^*_y)\check{\sigma}_z.
\label{two-level}
\end{equation}
$\check{\sigma}_{x,y,z}$ are the Pauli matrices in the 
basis of localized states $\ket{r}$ and $\ket{r^\prime}$;
and the $k$-independent constants are $\Delta_{\ell}=(-1)^{\ell}\Pi_{R=r+1}^{r^\prime-1}\left\{\frac{1}{2}\left[E_{r}(k^*_y)+E_{r^\prime}(k^*_y)\right]-E_R(k^*_y)\right\}^{-1}$ and $v_{\ell}=2\lambda\left[ \sin(2\pi\alpha r+ k^*_y)-\sin(2\pi\alpha r^\prime +k^*_y)\right]$. The eigenstates are ``Chern-dimers'', superpositions of $\ket{r}$ and $\ket{r^\prime}$ with wave function
\begin{equation}
\psi^\pm_{s_x}=\frac{1}{\sqrt{2}}\left(\delta_{s_x,r}+e^{i\beta_\pm}\delta_{s_x,r^\prime}\right),
\label{dimerwf}
\end{equation}
where $\beta_{+}=-\ell(k_x +\pi)+\pi$ and $\beta_{-} = -\ell(k_x +\pi)$ are relative phases for the upper and lower band edges, respectively.  Starting from Eq.~(\ref{two-level}) it is straightforward to show~\cite{fradkin91} that 
the Chern number of each band gap is $c_r=\ell$. Thus, for sufficiently anisotropic hopping, dimerized states form at the
band edges with spatial extent along ${\bf e}_x$ equal to the Chern number of the corresponding gap. 

The formation of dimers of size $\ell=c_r$ at the edges of the $r$-th gap implies 
a ``hidden spatial correlation'' described by correlation function 
$C(l) =\sum_{\mathbf{s}}\langle f^\dagger_{\mathbf{s}} f_{\mathbf{s}+l\mathbf{e}_x}\rangle$, 
which is peaked at $l=\ell$.  For chemical potential in the $r$-th gap,
$C(l)$ asymptotes to a delta function $C(l)\rightarrow \delta(l-\ell)$,
as $\lambda\rightarrow \infty$.  Because $\beta_+-\beta_- = \pi$ for each crossing, the net contribution to $C(l)$ from dimerized states associated with gaps fully below the Fermi energy is zero.  Since $C(l)$ and ${n}(k_x)$ are a Fourier transform pair, ${n}(k_x) = \sum_{l} C(l) \cos(k_x l)/\sqrt{N}$; this yields the $\cos(c_r k_x)$ term in Eq.~(\ref{nkx-asymp}).  The $r$ offset in Eq.~(\ref{nkx-asymp}) comes from
the $C(l=0)$ contribution. 

\section{Effect of temperature and trap}
\begin{figure}[b]
{\includegraphics[width =3.3in]{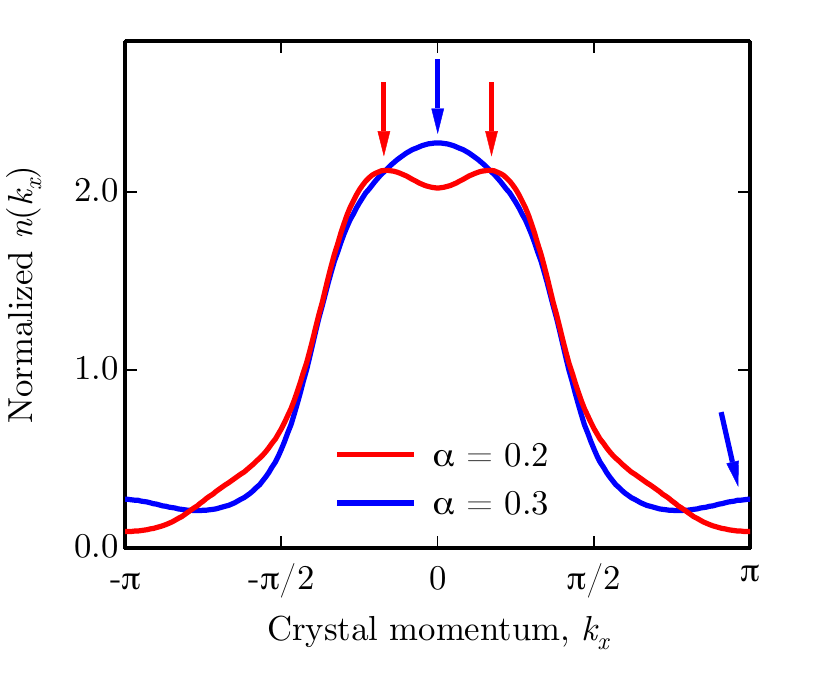}}
\leavevmode
\caption{(Color) The momentum distribution function $n(k_x)$ for the isotropic lattice 
at temperature $T=0.2t$ and a trap potential with $m\omega^2 d^2=10^{-3}t$. $\mu_0$ at the trap 
center is set to $-0.75t,-0.5t$ for
$\alpha=0.2,0.3$ respectively.
The twin peaks, indicated by the arrows, allow to determine the Chern
number near the trap center to be 2. The unit for $k_x$ is $1/d$.
}
\label{F4}
\end{figure}

Our numerical simulation based on Eq. (1) confirms that the structures in $n(k_x)$
persist, and remain visible, for trapped systems at finite temperatures. 
Due to the finite band gap of the quantum Hall insulator, thermal fluctuations only slightly smear the peak (or oscillation) structure in $n(k_x)$ as long as $T$ is small compared to the gap of interest, which is of order $t/2$ for the major gaps in
the Hofstadter spectrum. Fig. \ref{F4} shows examples of $n(k_x)$
for realistic trap and temperature parameters. Thus, our proposal
of detecting the Chern number is experimentally feasible.

\section{Q-matrix and momentum distribution}
Finally, we point out a general connection 
between $\tilde{n}(\mathbf{k})$ and the $Q$-matrix~\cite{ryu10},
which plays a fundamental role in the classification of topological insulators
and in turn is related to the Chern number of 
a quantum Hall insulator.
For a band insulator with band structure $E_r(\mathbf{k})$ 
(with energy measured from the chemical potential), the spectral projection
operators $P_+(\mathbf{k})$ and $P_-(\mathbf{k})$ project onto states above or below the chemical potential,
respectively. The $Q$-operator $Q(\mathbf{k})=P_+(\mathbf{k})-P_-(\mathbf{k})$
describes a ``flattened'' Hamiltonian that
is smoothly deformed from the original~\cite{ryu10}. 
The momentum distribution is related to the spectral projection operator of 
the filled bands $\rho(\mathbf{k})= P_-(\mathbf{k})$. In matrix form, $\rho$ is related to $Q$ by
\begin{equation}
Q_{ab}({\mathbf{k}})= \delta_{ab}-2\rho_{ab}({\mathbf{k}}).
\label{q}
\end{equation}
As an example, for the Hofstadter model at flux $\alpha=p/q$, 
$\rho_{ab}({\mathbf{k}})=\langle f_a^\dagger({\mathbf{k}}) f_b({\mathbf{k}})\rangle$, where
$a,b\in\left\{1,\ldots, q\right\}$ are 
the site indices within the magnetic unit cell, and ${\mathbf{k}}$ is the crystal momentum within
the MBZ. $\tilde{n}(\mathbf{k})$ as defined in Eq.~(\ref{MomentumDistribution}) is simply
$\tilde{n}(\mathbf{k})=\sum_{ab}\rho_{ab}(\mathbf{k})$. By Eq~(\ref{q}), for given $k_y=k^*_y$,
the behavior of $\tilde{n}(k_x,k^*_y)$ as function of $k_x$ is determined by 
the $Q$-matrix along cut $k_y=k^*_y$ in the BZ. So is
the 1D momentum distribution $n(k_x)$. On the other hand, the Chern number is
nothing but the winding number of the off-diagonal block of the $Q$-matrix in chiral basis,
also known as the transition function, along a cut that 
splits the BZ~\cite{ryu10}. Thus, on general grounds, phase winding responsible
for nonzero Chern number implies nontrivial structures in the momentum distribution function.
Explicitly, for the asymptotically exact effective Hamiltonian $\check{H}_{2}(\mathbf{k})$, 
the $Q$-matrix is $
\check{Q}(\mathbf{k})=(v_{\ell}^2k_y^2+\Delta_{\ell}^2)^{-1/2}\check{H}_2(\mathbf{k})$.
The Chern number $c_r = \ell$, for the off-diagonal block of $Q$-matrix is simply $\sim e^{ik_x\ell}$.
Meanwhile, from the $Q$-matrix, we obtain $n(k_x)=q\left[1-\mathrm{sign}(\Delta_\ell)
\cos ( k_x\ell)\right]$. This proves analytically our previous result Eq.~(\ref{nkx-asymp}). 
The oscillations adiabatically evolve into local peaks for the isotropic lattice, 
as explicitly shown in Fig.~\ref{F2}c and \ref{F2}d.  

In contrast with Chern numbers, $n(k_x)$ is not a topologically invariant quantity.  
Still, the clear signatures of Chern numbers in TOF images that we find here makes momentum distributions into an unexpected tool for exploring topological states of matter.  In addition to edge states and Hall conductance, ripples in the momentum distribution and dimerized states of anisotropic lattices whose spatial extent encodes the topological quantum numbers provide a new way to characterize and visualize non-trivial topological states.  Our study opens the possibility of using momentum distributions to explore other topological states of matter including topological superfluids.

We acknowledge the support of the ONR and NIST (E.Z. and I.I.S), the NRC (N.B.-A.), the NSF through the PFC at JQI (I.B.S.) and the ARO with funds from both the Atomtronics MURI and the DARPA OLE Program (I.B.S).
\bibliography{chern_1}

\end{document}